\documentclass[twocolumn,showpacs,amsmath,amssymb,aps]{revtex4-1}

\usepackage{graphicx}% Include figure files
\usepackage{dcolumn}% Align table columns on decimal point
\usepackage{bm}% bold math

\newcommand{\eps}{\epsilon}

\begin{document}

\preprint{APS/123-QED}

\title{Highly Dispersive Electron Relaxation and Colossal Thermoelectricity \\ in the Correlated Semiconductor FeSb$_2$}

\author{Peijie Sun$^1$}
\author{Wenhu Xu$^2$}
\author{Jan M. Tomczak$^3$}
\author{Gabriel Kotliar$^2$}
\author{Martin S\o ndergaard$^4$}
\author{Bo B. Iversen$^4$}
\author{Frank Steglich$^5$}
\affiliation{%
$^1$Beijing National Laboratory for Condensed Matter Physics, Institute of Physics, Chinese Academy of Sciences, Beijing 100190, China \\
$^2$Department of Physics and Astronomy, Rutgers University, Piscataway, New Jersey 08854, USA \\
$^3$Institute of Solid State Physics, Vienna University of Technology, A-1040 Vienna, Austria \\
$^4$Department of Chemistry, University of Aarhus, %Langelandsgade 140,
DK-8000 Aarhus C, Denmark \\
$^5$Max Planck Institute for Chemical Physics of Solids, 01187 Dresden, Germany
}%

\date{\today}

\begin{abstract}
  We show that the colossal thermoelectric power, $S(T)$, observed in the correlated semiconductor FeSb$_2$ below 30\,K is accompanied by a huge Nernst coefficient $\nu(T)$ and magnetoresistance MR$(T)$. Markedly, the latter two quantities are enhanced in a strikingly similar manner. While in the same temperature range, $S(T)$ of the reference compound FeAs$_2$, which has a seven-times larger energy gap, amounts to nearly half of that of FeSb$_2$, its $\nu(T)$ and MR$(T)$ are intrinsically different to FeSb$_2$: they are smaller by two orders of magnitude and have no common features. Underlying the essentially different thermoelectric properties between FeSb$_2$ and FeAs$_2$, a large mismatch between the electrical and thermal Hall mobilities was found only in the former compound. With the charge transport of FeAs$_2$ successfully captured by the density functional theory, we emphasize a significantly dispersive electron-relaxation time $\tau(\epsilon_k)$ related to electron-electron correlations to be at the heart of the peculiar thermoelectricity and magnetoresistance of FeSb$_2$.
\end{abstract}

\pacs{Valid PACS appear here}% PACS, the Physics and Astronomy
\maketitle

\section{introduction}

Current thermoelectric materials all operate around or above room temperature (RT) \cite{snyder08}. Efficient thermoelectric conversion at low temperatures ($T$\,$<$\,100\,K) by using a cryogenic Peltier cooler in, for example, microelectronic superconducting devices based on high-$T_c$ superconductors, may bring about a world-wide implementation of relevant technology \cite{mahan97}. Such thermoelectrics have not yet been discovered despite intensive efforts. Consequently, the recent observation of colossal thermoelectric power (TEP) in FeSb$_2$ amounting to tens of mV/K \cite{bentien07} at $T$\,$\approx$\,10\,K has been receiving considerable attention.

FeSb$_2$ is a new, $d$-electron based correlated semiconductor closely resembling FeSi \cite{petrovic05,luk06,sun11}. With the origin of the enhanced TEP of FeSi in debate for decades \cite{wolfe65,sales11,tomczak13}, the observations in FeSb$_2$, with the thermoelectric power factor the largest ever reported \cite{bentien07,martin13,jie12}, have added timely interest to this concern. Accumulating experimental facts have confirmed the significant effect of electron-electron correlations in FeSb$_2$ \cite{bentien07,petrovic05,luk06,sun11,perucchi,herzog10,martin13,jie12,hu09,sun09,sun10,hu12}. These include the large spectral weight redistribution up to higher energies \cite{perucchi,herzog10} and the occurrence of electronic Griffiths phases at low temperatures \cite{hu12}. The sign of the colossal TEP in FeSb$_2$ is negative, in agreement with the dominating electron band as revealed by Hall-effect measurements in the same window, $T$\,$<$\,30\,K (cf.\,Fig.\,1, inset (a)). However,  conventional  electronic  structure calculations  \cite{tomczak10} can only qualitatively produce the TEP profile, with the values being only one tenth of the measured ones.

In this work, we report on the Nernst coefficient $\nu(T)$, TEP $S(T)$ and magnetoresistance (MR) of FeSb$_2$ in comparison to the isostructural compound FeAs$_2$. The latter compound does not exhibit significant effects of electron correlations or is only weakly correlated \cite{sun10,tomczak10}. In FeSb$_2$, the simultaneously enhanced values of $S(T)$, $\nu(T)$ and MR$(T)$, as well as the striking similarity between the latter two quantities, evidence, as we will argue, a highly dispersive relaxation time $\tau(\epsilon_k)$ of the charge carriers to be at the heart of these enhanced quantities. Further support to this argument comes from the significant mismatch between the electrical and thermal Hall mobilities of FeSb$_2$, which forms a basis for our theoretical interpretation of the enhanced Nernst signal using a dispersive $\tau(\epsilon_k)$.  By contrast, for FeAs$_2$, $\nu(T)$ and MR$(T)$ show values which are two orders of magnitude smaller, with its TEP being well captured by the DFT approach assuming an energy-independent $\tau$ \cite{tomczak10}.

\section{Experimental Methods}

Single crystals of FeSb$_2$ and FeAs$_2$ were prepared by self-flux and chemical vapor transport techniques, respectively. Details of sample synthesis and structure characterization have been described elsewhere \cite{bentien07,sun10}.  The sample of FeSb$_2$ employed in this work is a 4.6$\times$2$\times$0.3\,mm$^3$ slab cut along the $c$ axis of the Marcasite structure, after orientation by the Laue diffraction. We stress that this particular sample has been used for all the transport measurements discussed in the present paper.
For the measurements of thermal properties including thermoelectric power, Nernst effect and thermal conductivity, we employed a home-made setup equipped with one chip resistor of 2000 $\Omega$ as heater and one thin ($\phi$ $=$ 25 $\mu$m) chromel-AuFe$_{0.07\%}$ thermocouple for detecting the temperature gradient \cite{urike}. Electrical resistivity and Hall-effect measurements were performed on the physical property measurement system (PPMS) from Quantum Design.

\section{Sample Dependence}

While a significant sample dependence is not unusual for a narrow-gap semiconductor, the one observed in the transport properties of FeSb$_2$ is surprisingly large. For example, different to the purely semiconducting behavior observed along all the crystallographic axes by us, Petrovic and coauthors \cite{jie12} reported that FeSb$_2$ crystals may show a metal-insulator transition along the $c$-axis, depending on the conditions of decanting after the flux growth.

Here we focus particularly on the sample dependence of thermoelectric power. The single crystals of FeSb$_2$ prepared by us exhibit a colossal TEP minimum at around 10 K, which, depending on the sample quality, amounts to a value between -10 and -45mV/K \cite{bentien07,sun10}. Due to the high thermal conductivity which amounts to about 500 W/mK at around 10 K, thermal measurements on this compound require to pay very careful attention on thermal stability. For this reason, we used a very small temperature gradient during the thermal measurements which was typically 0.002$\times$$T$, with $T$ being the sample base temperature. We have confirmed these minimum values of TEP by different measurement techniques that include the original and modified thermal transport options on PPMS (cf. ref. \cite{bentien07} and its Supplementary Material), as well as the home-made setup employed in this work. We recognized that the TEP minima reported by the groups of Petrovic and Takahashi, which are obtained on the original PPMS platform, are roughly one order of magnitude smaller (1$-$2 mV/K) \cite{jie12,kefeng12,takahashi11}. Nevertheless, in these cases, the thermoelectric power factor, $S^2/\rho$, could be much larger than our observations due to a reduced electrical resistivity \cite{jie12}. It is unfortunate that no thermal conductivity results are available for those FeSb$_2$ samples \cite{jie12,takahashi11}, except for the sample in ref. \cite{kefeng12}, which exhibits a metal-insulator transition and therefore prevents a direct comparison with our thermal conductivity data.

Another interesting point is that the smaller TEP minima of FeSb$_2$ single crystals in the literature were observed either to span over a large temperature region between 10 and 20 K \cite{jie12}, or at a rather higher temperature, 20 K \cite{takahashi11,kefeng12}, in comparison to 10 K as observed by us. This may hint at a better quality of the samples used in this work. For example, a tiny amount of electron doping by Te (0.05\%) into the Sb sites can shift the TEP minimum to 30 K, with its magnitude reduced by more than one order of magnitude \cite{sunapl11}.  In support of the peculiar thermal transport in FeSb$_2$, recently, in the isostructural semiconductor CrSb$_2$, a huge TEP minimum of $-$4.5 mV/K at 18 K and thermal conductivity comparable with that of FeSb$_2$ were reported by Sales $et$ $al.$ \cite{sales12}.

\section{results and discussion}

An increase of the electrical resistivity, $\rho(T)$, by 4 orders of magnitude upon cooling from RT to 2\,K evidences a high quality of the current FeSb$_2$ crystal (cf.\,Fig.\,1). As had been reported for various FeSb$_2$ samples \cite{bentien07,jie12,sun10}, a shoulder in $\rho(T)$ is observed between 10 and 20\,K. Applying the thermal activation law to $\rho(T)$ between 40 and 100\,K yields an energy gap $E_g$\,$\simeq$\,28\,meV, which can be reasonably explained by the electronic structure calculations only when taking electron-electron correlations into account through, e.g., Hedin's GW approximation \cite{tomczak10}. At even lower temperatures ($T$\,$<$\,15\,K), $\rho(T)$ is found to be dominated by a small activation energy of  $\sim$\,6\,meV. These transport features were found to be rather robust against varying sample quality and crystallinity, except for a sample-dependent activation energy ($<$\,0.1\,meV) which can be estimated from the nearly flat $\rho(T)$ below 6\,K. The transport properties in this low-temperature region ($T$ $<$ 6 K) are assumed to be influenced by impurity bands and are therefore excluded from our focus in this work.
For FeAs$_2$, thermal activation behavior of $\rho(T)$ was confirmed above 200\,K (Fig.\,1), with $E_g$\,$\sim$\,200\,meV. The metallic behavior between 50 and 180\,K in this compound presumably is caused by in-gap impurity states.

\begin{figure}
\includegraphics[width=0.95\linewidth]{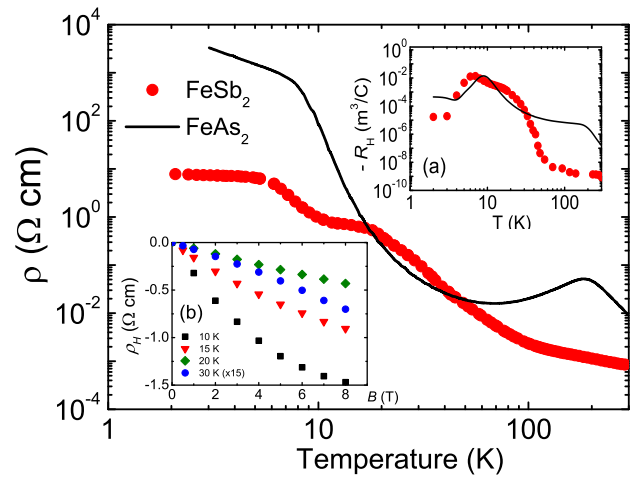}
\caption{Electrical resistivity $\rho(T)$ (main panel), and Hall coefficient $R_H(T)$ (inset\,(a)) for FeSb$_2$ and FeAs$_2$. $R_H(T)$ was measured at $B$\,$=$\,1\,T. Inset (b): isothermal Hall resistivity $\rho_H(B)$ at selected temperatures for FeSb$_2$.
\label{nu.eps}}
\end{figure}

It is known that the Hall coefficient $R_H(T)$ of FeSb$_2$ is subject to multi-band competition \cite{jie12,sun10,2takahashi11,hu08,takahashi13}.
This effect, however, is largely sample dependent: at $T$\,$<$\,30\,K, besides a high-mobility electron band as reported in the literature, the additional low-mobility band could be either hole-like \cite{jie12,hu08} or electron-like (this work and refs. \cite{2takahashi11,takahashi13}). A clear distinction between the two cases with opposite low-mobility carriers is reflected by the opposite curvature of the Hall resistivity $\rho_H(B)$ (cf., Fig.\,1 inset\,(b) and inset\,(a) of Fig.\,2 in ref.\,\cite{jie12}). In spite of the significant sample dependence concerning the low-mobility band, it has been concluded that in the temperature window of interest, i.e., 6$-$30\,K, the electrical transport of FeSb$_2$ is dominated by the high-mobility electron band \cite{jie12,2takahashi11,takahashi13}, in line with the sublinear $\rho_H(B)$ curves as shown in the inset\,(b) of Fig.\,1. Upon warming the compound up to above 40\,K, while we find the high-mobility band still to be of electron character, it could be hole-like in other samples \cite{jie12,hu08}. Given a dominating one-band transport in the FeAs$_2$ sample \cite{sun10}, the two compounds investigated here have a similar carrier concentration below 30\,K (cf. Fig.\,1 inset (a)), in which temperature range the TEP assumes a peak in either case.

\begin{figure}
\includegraphics[width=0.9\linewidth]{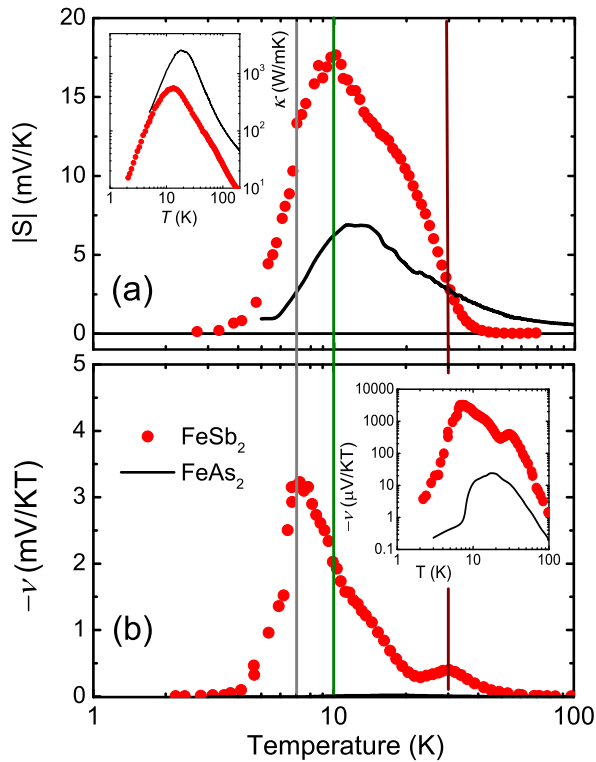}
\caption{The thermoelectric power $S(T)$ (a) and Nernst coefficient $\nu(T)$ (b) for FeSb$_2$ and FeAs$_2$. Inset of (a) shows the thermal conductivity $\kappa(T)$ of these two compounds. Note that $\nu(T)$ of FeAs$_2$ is smaller than that of FeSb$_2$ by two orders of magnitude, therefore being almost hidden in the horizontal axis of (b). For clarity, the same data are also displayed in a double-log plot in the inset of (b).
\label{nus.eps}}
\end{figure}

As shown in Fig.\,2(a), upon cooling the TEP of the current FeSb$_2$ sample starts to be enhanced from 40\,K and assumes a maximum absolute value of 17.5\,mV/K at $T$\,$\approx$\,10\,K. In the same temperature window, the Nernst coefficient $\nu(T)$, measured at $B$\,$=$\,2\,T, is strongly enhanced as well. It shows two peaks at 7\,K and 30\,K, as well as an anomaly at 10\,K, i.e., the position where the TEP peak occurs (Fig.\,2(b)). A similar $\nu(T)$ curve with double-peak profile has already been observed in another FeSb$_2$ crystal \cite{sun09}. All these anomalies in the $\nu(T)$ curve can actually find their counterparts in $S(T)$, e.g., the main peak of $\nu(T)$ at 7\,K concurs with a jump in $S(T)$, as indicated by the gray vertical line. Further on, the multiple anomalies are not only limited to $\nu(T)$ and $S(T)$, but clearly show up in the MR$(T)$ (cf. Fig.\,3) and the d$\rho$/d$T$ vs $T$ curves \cite{jie12} as well. We wish to particularly stress that the $\nu(T)$ and the MR$(T)$ curves of FeSb$_2$ are surprisingly similar concerning their complex temperature profiles. This, as will be discussed below, strongly suggests a significantly dispersive $\tau(\epsilon_k)$ of charge carriers.

\begin{figure}[t]
\includegraphics[width=0.95\linewidth]{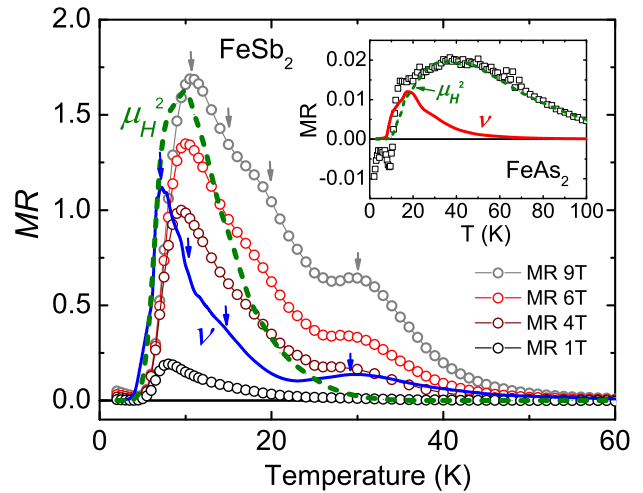}
\caption{Magnetoresistance MR\,$=$\,$(\rho_B\,-\,\rho_0)/\rho_0$ for FeSb$_2$ measured at various magnetic fields (main panel), and for FeAs$_2$ at
4\,T (inset). Note that the MR$(T)$ of the latter compound is smaller by a factor of 50 than that of FeSb$_2$. For comparison, the Nernst coefficient $\nu(T)$ and the square of the electrical Hall mobility, $\mu_H^2(T)$, are shown in arbitrary units. While for FeSb$_2$, $\nu(T)$ reveals a strikingly similar $T$-profile compared to MR$(T)$, as indicated by arrows on both $\nu(T)$ and MR$_{9\,T}$$(T)$ curves, for FeAs$_2$, by contrast, MR$(T)$ scales well with $\mu_H^2(T)$.
\label{nus.eps}}
\end{figure}

The huge Nernst coefficient of FeSb$_2$ with a maximum of 3.2\,mV/KT at 7\,K  dwarfs that of FeAs$_2$ (Fig.\,2(b) main panel and inset) by two orders of magnitude. This difference of $\nu(T)$ in magnitude is extremely prominent when compared to that in TEP: $S(T)$ of FeAs$_2$ assumes a peak below 20\,K amounting to 7\,mV/K, i.e., a value as large as 40\% of that in FeSb$_2$. As we will discuss below in terms of $\nu$/$\mu_H$, where $\mu_H$ is the Hall mobility (cf. Fig.\,4), even if the largely differing Hall mobilities of the two materials are taken into account, such a large difference in $\nu(T)$ is surprising. On the other hand, the large values of TEP in FeAs$_2$ are well captured by DFT \cite{tomczak10}, which predicts a reasonable charge gap of $\sim$\,0.2\,eV as observed experimentally \cite{sun10}. Because of the much smaller charge gap ($\sim$\,28\,meV), the colossal TEP observed in FeSb$_2$ is far beyond the upper bound ($\sim$\,1.5\,mV/K) set up in the coherent electron diffusion picture\cite{tomczak10}.

A dominant phonon-drag contribution has been frequently assumed in order to explain the enhanced TEP in FeSb$_2$ \cite{tomczak10,takahashi11,2takahashi11}. However, this is doubtful because (i) in FeAs$_2$, while the thermal conductivity $\kappa(T)$ is larger than that of FeSb$_2$ by a factor of 10 (Fig.\,2(a), inset), compatible with an enhanced phonon-drag contribution, $S(T)$ can nevertheless be reasonably explained within the DFT frame; (ii) the phonon-drag effect may contribute to the Nernst effect as well, as observed, e.g., in the semimetal Bi \cite{behnia07}. In FeSb$_2$, however, the complex profile of $\nu(T)$ shows striking resemblance to that of MR($T$), instead of $\kappa(T)$ as is usually expected for dominating phonon-drag thermoelectric transport. Since MR$(T)$ is free of phononic transport, our observations in FeSb$_2$ strongly argue against a phonon-drag scenario, providing evidence for both $\nu(T)$ and MR$(T)$ being dominated by a common electronic origin.

\begin{figure}[t]
\includegraphics[width=0.95\linewidth]{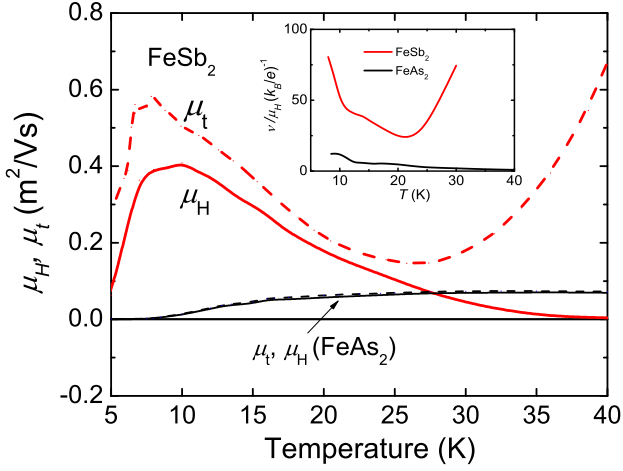}
\caption{Comparison of the electrical and thermal Hall mobilities, $\mu_H(T)$ and $\mu_t(T)$, for FeSb$_2$ and FeAs$_2$.  Note that $\mu_H(T)$ and $\mu_t(T)$ are significantly different only in the former compound. Inset: dimensionless ratio $\nu$/$\mu_H$($k_B/e$)$^{-1}$ for the two compounds.
\label{nus.eps}}
\end{figure}

In terms of the linear response coefficients, the charge current {\bf J} can be induced by external electric field {\bf E} or thermal gradient $\nabla T$:
${\bf J}=\bar{\sigma}\cdot{\bf E}-\bar{\alpha}\cdot\nabla T$.
We consider {\bf E}, $\nabla T$ and thus {\bf J} to be in the $x$-$y$ plane, hence the two conductivities $\bar{\sigma}$ (electrical) and $\bar{\alpha}$ (Peltier) are 2$\times$2 matrices with diagonal and off-diagonal elements in the presence of a magnetic field $B$ along the $z$ axis.
The Nernst effect measures the transverse electric field $E_y$ and can be rooted to the difference of two mobilities (for more information, refer to the Appendix),
\begin{equation}
\begin{aligned}
\nu\,&\equiv\,\frac{E_y}{B\nabla_xT}\,=\,\frac{1}{B}\frac{\alpha^{yx}\sigma^{xx}-\alpha^{xx}\sigma^{yx}}{\sigma^{xx}\sigma^{yy}+\sigma^{yx}\sigma^{xy}}\\
&\simeq\,\frac{\alpha^{xx}}{\sigma^{xx}}\left(\frac{1}{B}\frac{\alpha^{yx}}{\alpha^{xx}}-\frac{1}{B}\frac{\sigma^{yx}}{\sigma^{xx}}\right)\\
&=\, S(\mu_t-\mu_H),
\end{aligned}
\end{equation}
where $S$\,=\,$\frac{\alpha^{xx}}{\sigma^{xx}}$ defines the TEP, $\mu_H$\,=\,$\frac{1}{B}\frac{\sigma^{yx}}{\sigma^{xx}}$ is the Hall mobility and can be simply estimated by computing $R_H/\rho$. We furthermore define $\mu_t$\,=\,$\frac{1}{B}\frac{\alpha^{yx}}{\alpha^{xx}}$  as the thermal analogue of $\mu_H$.

The electrical and thermal Hall mobilities, $\mu_H$ and $\mu_t$, as derived from our measured data presented in Figs.\,1-2 and Eq.\,1 are shown in Fig.\,4 for both FeSb$_2$ and FeAs$_2$. For the latter compound, $\mu_t$ and $\mu_H$ are practically on top of each other and approximately constant over a large temperature range. By contrast, there is a significant difference between these two quantities for FeSb$_2$. Tracing the enhanced Nernst signal in FeSb$_2$ to the mismatch in the two mobilities, in our opinion, is an important step forward in the understanding of this material.

The divergent behavior of $\mu_t$ as well as the rapid decrease of $\mu_H$ above 30\,K for FeSb$_2$ are partly ascribed to an ambipolar effect in the presence of two carrier bands of opposite signs. This situation is supported by Hall-resistivity measurements: above 30 K, $\rho_H(B)$ changes its curvature (cf. Fig.\,1, inset (b)), indicating that the low-mobility band changes from electron to hole like (note that the high-mobility band remains electron like).
 As discussed by Bel $et$ $al.$ \cite{behnia03}, in a compensated two-band system,
\begin{equation}
\mu_t=\frac{1}{B}\frac{\alpha_+^{yx}+\alpha_-^{yx}}{\alpha_+^{xx}+\alpha_-^{xx}};\,\,\mu_H=\frac{1}{B}\frac{\sigma_+^{yx}+\sigma_-^{yx}}{\sigma_+^{xx}+\sigma_-^{xx}},
\end{equation}
where the subscript denotes the sign of the carriers in the respective band. Following the explanation in ref.\,\cite{behnia03}, one readily knows that while $\sigma_+^{xx}$ and $\sigma_-^{xx}$ have the same sign, $\sigma_+^{yx}$ and $\sigma_-^{yx}$ are opposite in sign, as is also evident from the opposite Hall signals of $+$ and $-$ bands. On the other hand, the Peltier conductivity $\alpha_+^{xx}$ and $\alpha_-^{xx}$ (therefore $S$ of the two bands) have opposite signs, whereas $\alpha_+^{yx}$ and $\alpha_-^{yx}$ have the same sign. Therefore these combinations lead to an enhancement of $\mu_t$ and a decrease of $\mu_H$ above 30 \,K. Our following discussions apply only to the lower temperature range (6 $-$ 30 K) where we assume a one-band approximation to be adequate.

In order to demonstrate the role of a dispersive $\tau(\epsilon_k)$ in various transport coefficients, we employ a model of conventional semiconductor, where the chemical potential $\mu$ = $-$$\Delta$, with $\Delta$ being the activation energy. The conduction band is approximated in terms free electrons with an effective mass $m^*$, $\epsilon_k$\,$=$\,$\frac{\hbar^2k^2}{2m^*}$. This dispersion is supplemented with a scattering rate that has a linear energy dependence, $\Gamma_k$\,$\simeq$\,$\Gamma_0(T)$\,+\,$a(T)$$\epsilon_k$, where $\Gamma_0(T)$ is the scattering rate of charge carriers at the bottom of the conduction band, and $a(T)$ is a dimensionless parameter. Note that the linear energy dependence is the simplest dispersion of $\Gamma_k$, which is employed here in order to demonstrate the impact of an asymmetry of $\Gamma_k$ on the transport properties, rather than to capture the physics in FeSb$_2$.
In the Boltzmann framework, we have shown that all components of the conductivities $\bar{\sigma}$ and $\bar{\alpha}$ (cf.\,Eq.\,1) can be expressed
in terms of $m^*$, $\frac{\triangle}{k_BT}$, $\frac{\Gamma_0}{k_BT}$, and $a(T)$ (refer to the Appendix).

As shown in the Appendix, when $a(T)$\,$=$ 0, i.e., $\Gamma_k$ being energy independent, the mobilities $\mu_t$ and $\mu_H$ cancel exactly, generalizing the Sondheimer cancellation theorem of the Nernst effect. This is consistent with the analytical explanation made for typical metals, where $\nu$\,$\propto$ $\partial$$\tau(\epsilon _k)$/$\partial \epsilon _k$ \cite{behnia09}. The nonvanishing, and in fact large, Nernst coefficient (or, more precisely, the large mismatch between $\mu_t$ and $\mu_H$) in FeSb$_2$ indicates that a strong energy-dependent scattering rate plays an important role in the thermoelectricity of this compound. This is also evident in the large values of MR$(T)$ in FeSb$_2$, in comparison to FeAs$_2$ (cf. Fig.\,3): In the formalism as employed above, the MR can be expressed with the same set of parameters, and it vanishes as well when $a(T)$\,=\,0.

Bearing in mind our proposition of a common underlying origin of $\nu$ and MR (cf. the Appendix), it is interesting to note the striking similarities observed between these two quantities for FeSb$_2$ as a function of temperature, as revealed by Fig.\,3. To our knowledge, this phenomenon has never been reported for any other compound. Qualitatively similar MR$(T)$ with double peaks was also observed by Hu $et$ $al.$ \cite{hu08mr}. There, however, the peak at the high-$T$ side ($\sim$\,30\,K) is nearly two orders of magnitude larger than that of our observations, compatible with the metal-insulator transition of their samples.
A slight shift of the double peaks in MR$(T)$ to higher temperatures upon increasing field is observed in Fig.\,3, indicating the involvement of the spin degrees of freedom in the relevant charge-carrier relaxation processes. Notice, however, that even a large value of $a(T)$ is not enough to account quantitatively for the observed $\nu(T)$ of FeSb$_2$, suggesting that the assumed linear energy dependence of $\Gamma_k$ cannot correctly capture the correlation effects in FeSb$_2$. A more complicated dispersion of $\Gamma_k$ has to be considered in order to make comparison with the experimental results. In addition, impurity states and/or phonon-drag effect which could lead to a difference between $\mu_t$ and $\mu_H$ may also play a role to some extent.

As shown in the inset of Fig.\,3, the MR$(T)$ of FeAs$_2$ measured at 4\,T spans vertically a range which is a factor-of-50 smaller than that for FeSb$_2$. This is in line with the values of its $\nu(T)$ which are roughly two orders of magnitude smaller than those of FeSb$_2$. Different to FeSb$_2$, no obvious correlation can be observed between $\nu(T)$ and MR$(T)$ for this compound. Instead, we found that the MR($T$) curve scales well with the $\mu_H^2(T)$ curve in FeAs$_2$, as expected for a conventional semiconductor \cite{Seeger}.

The energy-dependent scattering processes in FeSb$_2$ can also be highlighted by simply computing the ratio $\nu$/$\mu_H$ in units of $k_B/e$. This is because in a nondegenerate semiconductor, $\nu\,=\,-\frac{k_B}{e} \mu _H r$ \cite{Seeger}, if $\tau(\epsilon_k)$ obeys a power law function, $\tau(\epsilon_k)$\,$\propto$\,$\epsilon_k ^r$. A similar scaling between Nernst coefficient and Hall mobility has indeed been demonstrated for various metallic systems in ref. \cite{behnia09}, where $\tau(\epsilon_k)$ has been assumed to obey a linear function of energy, i.e., $r$ = 1.
As shown in the inset of Fig.\,4, the ratio $\nu$/$\mu_H$ in units of $k_B/e$, as a measure of the power $r$, is unrealistically large for FeSb$_2$. Whereas in FeAs$_2$, it is of order unity as anticipated for, e.g., acoustic-phonon or ionized-impurity scattering \cite{Seeger}. This difference, again, hints at a highly energy-dependent scattering rate in FeSb$_2$ that is presumably related to the substantial electron-electron correlations in this compound. This may be a local resonant scattering similar to that operating in Kondo systems \cite{sun12}, or incoherent electronic transport beyond the Landau quasiparticle picture as realized in doped Mott insulators \cite{gabi98}.

\section{concluding remarks}

The thermoelectric power of a conducting solid is related to its energy-dependent electrical conductivity \cite{mott}. Consequently, $S$ arises from the energy dependence of the two crucial components of electrical conductivity at the Fermi energy: the electronic density of states and the relaxation time $\tau(\epsilon_k)$. Among them, the latter effect is of minor importance for most materials \cite{snyder08}. However, we have recently shown for the prototype Kondo lattice compound CeCu$_2$Si$_2$ that its $\nu(T)$ and $S(T)$ are simply related by the electrical Hall mobility $\mu_H$ \cite{sun12}. Physically, this occurs only when the dispersive $\tau(\epsilon_k)$ is the dominating cause of the enhanced thermoelectric power as well as the Nernst coefficient, free of the Sondheimer cancellation that occurs when $\tau(\epsilon_k)$  is independent or only weakly dependent on energy.

Along the same line, a strongly energy-dependent scattering mechanism, as indicated for FeSb$_2$ by the largely and similarly enhanced Nernst coefficient and magnetoresistance, is supposed to play a significant role in the pronounced thermoelectricity. Theoretically, we propose to trace the enhancement of Nernst coefficient in FeSb$_2$ to the mismatch between the thermal and electrical Hall mobilities as observed in this material. We show that the mismatch of the two mobilities can stem from a dispersive scattering rate of the charge carriers, though our assumption of a linear energy dependence is too simple to capture the physics in FeSb$_2$. The unique scattering mechanism indicated for FeSb$_2$, which is absent in the uncorrelated band semiconductor FeAs$_2$, appears to be related to substantial electron-electron correlations, as manifested by a large (factor of 20$-$50) renormalization of the charge-carrier mass in slightly Te-doped FeSb$_2$ \cite{sun11,hu12}. A proper theory of FeSb$_2$ explaining this mass renormalization and the dispersive $\tau(\epsilon_k)$ has still to be developed.

\begin{acknowledgments}
This work is supported by the MOST of China Grant No. 2012CB921701 (P.S.), the NSF grant No. DMR-0906943 (W.X. and G.K.), the Danish National Research Foundation DNRF93 (M.S. and B.B.I.), and the DFG through FG 960 ``Quantum Phase Transitions" (F.S.).
\end{acknowledgments}

\appendix

\section{Theoretical Treatments of the Transport Conductivities}

We compute the Nernst coefficient $\nu(T)$ and magnetoresistance MR$(T)$ of a
single-band semiconductor within the framework of Boltzmann theory. We show
that if the charge carriers' scattering rate is non-dispersive, i.e.,
$\Gamma_k(\eps_k) = \Gamma_0$ is independent of the carriers' energy $\eps_k$, $\nu(T)$ and MR$(T)$ will
vanish. Also we show that while
a simple linear dispersion $\Gamma_k(\eps_k) = a\eps_k$ is able to produce a non-vanishing Nernst coefficient and magnetoresistance, it is not sufficient to explain the colossal Seebeck and Nernst coefficients of FeSb$_2$. This
indicates that the correlation effects in FeSb$_2$ cannot be correctly captured by such a simple dispersion of $\Gamma_k$.

Consider the charge current $\mathbf{J}$ as a response to an electric field
$\mathbf{E}$ and temperature gradient $\mathbf{\nabla} T$:
\begin{equation}
 \mathbf{J} = \bar{\sigma}\cdot \mathbf{E} - \bar{\alpha} \cdot \mathbf{\nabla} T.
\end{equation}
$\mathbf{J}$, $\mathbf{E}$, and $\mathbf{\nabla} T$ lie in the $xy$ plane, thus the two conductivities
$\bar{\sigma}$ and $\bar{\alpha}$ are $2 \times 2$ matrices.

A magnetic field $B$ perpendicular to the $xy$ plane gives rise to non-zero values of $\sigma^{xy}$ and $\alpha^{xy}$.
Besides, the longitudinal conductivities $\sigma^{xx}$ and $\alpha^{xx}$ also acquire $B$-dependent contribution, e.g.,
$\sigma^{xx} = \sigma^{xx}_0+\delta\sigma^{xx}(B)$. Hereafter the subscript $0$ denotes the absence of
magnetic field. The resistivity $\rho$, Hall mobility $\mu_H$, thermoelectric power
$S$, and magnetoresistance MR are defined as
\begin{eqnarray}
 \rho &\equiv& \frac{E_x}{J^x} = \frac{1}{\sigma^{xx} + \frac{\sigma^{xy}\sigma^{xy}}{\sigma^{xx}}}, \label{eq:rho} \\
 \mu_H &\equiv& \frac{E_y}{BE_x} = \frac{\sigma^{xy}}{B\sigma^{xx}}, \label{eq:mu_H} \\
 S &\equiv& \frac{E_x}{\mathbf{\nabla} T} = \frac{\alpha^{xx}}{\sigma^{xx}}, \label{eq:S} \\
  {\rm MR} &\equiv& \frac{\rho -\rho_{0}}{\rho_{0}}. \label{eq:MR}
\end{eqnarray}
$J^y=0$ has been used in writing Eqs.~\ref{eq:rho} --~\ref{eq:MR}. In a common setup of Hall measurement, the $y$-component
of $\mathbf{E}$ is established due to transverse motion ($J^y$) of carriers, which counteracts $J^y$ and
leads to $J^y=0$.

We are concerned with the leading order in $B$ for the transport quantities defined above.
When $B$ is small, $\sigma^{xy}$ and $\alpha^{xy}$ are proportional to $B$, and $\delta\sigma^{xx}$ is
proportional to $B^2$. Thus the leading orders are given by $\rho = 1/\sigma^{xx}_0$, $\mu_H = \sigma^{xy}/B\sigma^{xx}_0$,
and $S = \alpha^{xx}_0/\sigma^{xx}_0$. The subleading term of $\rho$ gives rise to MR. Hence,
\begin{eqnarray}
 \rho &=& \frac{1}{\sigma^{xx}_{0}+\left(\frac{\delta\sigma^{xx}}{B^2}+\sigma^{xx}_{0}\mu_H^2\right)B^2} \nonumber \\
 {}&\simeq& \rho_{0} - \left(\frac{\delta\sigma^{xx}}{B^2\left(\sigma^{xx}_{0}\right)^2}+\frac{\mu_H^2}{\sigma^{xx}_{0}}\right)B^2,
\end{eqnarray}
and
\begin{equation}
 {\rm MR} = -\left(\frac{\delta\sigma^{xx}}{B^2\sigma^{xx}_0} + \mu_H^2\right)B^2. \label{eq:MR_exp}
\end{equation}

The Nernst coefficient $\nu$ is the transverse counterpart of $S$,
\begin{eqnarray}
  \nu &\equiv& \frac{E_y}{B\nabla_x T} = \frac{1}{B}\frac{\alpha^{xy}\sigma^{xx}-\alpha^{xx}\sigma^{xy}}{\sigma^{xx}\sigma^{xx}+\sigma^{xy}\sigma^{xy}} \nonumber \\
  {}&\simeq& \frac{\alpha_0^{xx}}{\sigma_0^{xx}}\left(\frac{\alpha^{xy}}{B\alpha_0^{xx}}-\frac{\sigma^{xy}}{B\sigma_0^{xx}}\right) \nonumber \\
  {}& = & S(\mu_t - \mu_H).
\end{eqnarray}
We have defined a ``thermal'' counterpart of the Hall mobility, $\mu_t = \frac{\alpha^{xy}}{B\alpha_0^{xx}}$.

In Boltzmann theory, the conductivities are determined by the Fermi distribution function,
band velocity (and its higher order derivatives), and the relaxation time. To model a single-band
semiconductor of electron-type carriers, we assume a parabolic conduction band,
\begin{equation}
 \eps_k = \frac{\hbar^2 k^2}{2 m^*},
\end{equation}
where $m^*$ is the effective mass. The band velocity and higher order
derivative are written as
\begin{equation}
 v^{\alpha\beta\dots}_k = \frac{\partial \eps_k}{(\hbar \partial k_{\alpha})(\hbar \partial k_{\beta})\dots}.
\end{equation}
The Fermi distribution function is
\begin{equation}
 f(\eps) = \frac{1}{1+\exp\left(\frac{\eps_k+\Delta}{k_B T}\right)}.
\end{equation}
A fixed chemical potential,
$\mu = -\Delta$, is chosen. $\Delta$ is the activation energy. The relaxation time, $\tau_k$, of
a carrier with energy $\eps_k$ is determined by the scattering rate $\Gamma_k$,
\begin{equation}
 \tau_k = \frac{\hbar}{2\Gamma_k(\eps_k)}.
\end{equation}
Then the conductivities can be computed readily by the following expressions.
\begin{eqnarray}
 \sigma^{xx}_0 &=& \frac{2e^2}{V}
 \sum_k\left(-\frac{\partial f}{\partial \eps_k}\right)v^x_kv^x_k\tau_k.
 \label{eq:sigma_xx} \\
 \frac{\sigma^{xy}}{B} &=& \frac{2e^3}{V}\sum_k
 \left(-\frac{\partial f }{\partial\eps_k}\right)v^x_kv^x_kv^{yy}_k\tau_k^2,
 \label{eq:sigma_xy} \\
 \alpha^{xx}_0 &=& -\frac{2ek_B}{V}
 \sum_k\left(-\frac{\partial f}{\partial \eps_k}\right)v^x_kv^x_k\left(\eps_k+\Delta\right)\tau_k.
 \label{eq:alpha_xx} \\
 \frac{\alpha^{xy}}{B} &=& -\frac{2e^2 k_B}{V}\sum_k
 \left(-\frac{\partial f }{\partial\eps_k}\right)v^x_kv^x_kv^{yy}_k\left(\eps_k+\Delta\right)\tau_k^2, \nonumber \\
 \label{eq:alpha_xy} \\
 \frac{\delta\sigma^{xx}}{B^2} &=& \frac{2e^4}{V}\sum_k\left(-\frac{\partial f }{\partial\eps_k}\right)
 \left( v^x_kv^x_kv^{xx}_kv^{yy}_k \right. \nonumber \\
 {}&&-\left. v^{xxx}_kv^x_kv^y_kv^y_k \right)\tau_k^3 \label{eq:dsigma_xx}
 \end{eqnarray}

We assume the energy dependence of the scattering rate $\Gamma_k$ has
a non-dispersive part and a linear part,
\begin{equation}
 \Gamma_k \simeq \Gamma_0(T) + a(T)\eps_k.
\end{equation}
This form is often adopted in understanding the thermoelectric transport in
correlated metals.

If the carrier relaxation rate is non-dispersive, i.e., $a(T)$ = 0 and $\Gamma_k$ = $\Gamma_0$,
we can get the following conclusions straightforwardly for the non-degenerate limit, i.e., $\Delta \gg k_B T$,
\begin{eqnarray}
 \rho &\simeq& \left(\frac{2\pi}{k_B T}\right)^{3/2} \frac{\hbar^3}{e^2 (m^*)^{1/2} \tau_0}\exp\left(\frac{\Delta}{k_B T}\right), \label{eq:rho_gamma0} \\
 S &\simeq& -\frac{k_B}{e} \left(\frac{\Delta}{k_B T}+\frac{5}{2}\right). \label{eq:S_gamma0}
\end{eqnarray}
These are expected results for a conventional semiconductor. $\tau_0 = \hbar/2\Gamma_0$ is the non-dispersive
relaxation time.

The Hall mobility $\mu_H$ and its thermal counterpart $\mu_t$ are simply
\begin{equation}
 \mu_H = \mu_t = \frac{e\tau_0}{m^*}. \label{eq:mu_gamma0}
\end{equation}
Eq.~\ref{eq:mu_gamma0} does not require the non-degenerate limit, thus it is more general than
Eqs.~\ref{eq:rho_gamma0} and \ref{eq:S_gamma0}. Therefore, $\mu_H$ and $\mu_t$ exactly cancel out
 in the Nernst effect, $\nu = S(\mu_t-\mu_H)$, leading to a vanishing $\nu$.
Also, the two terms on the right hand side of Eq.~\ref{eq:MR_exp} cancel each other, leading
to vanishing magnetoresistance as well.

If the scattering rate is simply linear in energy, $\Gamma_k = a(T)\eps_k$,
we have following results in the non-degenerate limit,
\begin{eqnarray}
 \rho &\simeq& \left(\frac{9\pi^3}{2}\right)^{1/2} \left(\frac{\hbar}{e}\right)^2 \frac{a}{m^*k_B T} \exp\left(\frac{\Delta}{k_B T}\right), \label{eq:rho_a} \\
 S &\simeq& -\frac{k_B}{e} \left(\frac{\Delta}{k_B T} + \frac{3}{2}\right). \label{eq:S_a}
\end{eqnarray}
Thus the temperature dependence of $\rho$ and $S$ is dominated by the activation
energy $\Delta$, similar to the results of the non-dispersive relaxation.

The Hall mobility is
\begin{equation}
 \mu_H = \frac{\hbar e}{m^*k_B T} \frac{1}{a}. \label{eq:mu_H_a}
\end{equation}
The thermal mobility $\mu_t$ is
\begin{equation}
 \mu_t \simeq \frac{\hbar e}{m^*k_B T} \frac{1}{a} \left(1-\frac{k_B T}{\Delta}\right). \label{eq:mu_t_a}
\end{equation}
The $k_B T/\Delta$ term in Eq.~\ref{eq:mu_t_a} is the subleading order in the non-degenerate limit.
Therefore, $\mu_t$ and $\mu_H$ do not cancel and give rise to a non-vanishing $\nu$.
\begin{equation}
 \nu = \frac{k_B}{e}\mu_H \left(1+\frac{3}{2}\frac{k_B T}{\Delta}\right).
\end{equation}
But we emphasize that the difference between $\mu_H$ and $\mu_t$,
\begin{equation}
 \mu_t - \mu_H = -\frac{k_B T}{\Delta} \mu_H,
\end{equation}
should be small at the non-degenerate limit, and vanishes with
decreasing temperature. Therefore, our analysis shows that while a linear dispersion of $\Gamma_k$ is able to produce a finite Nernst signal, obviously, it is not consistent with the experimental results in FeSb$_2$, especially in the low temperature range where the colossal Seebeck
and Nernst coefficients are measured. Therefore, a complex but realistic dispersion of $\Gamma_k$ is yet to be identified for FeSb$_2$ in future.

Similarly, the magnetoresistance is
\begin{equation}
 \frac{MR}{B^2} \simeq \mu_H^2 \left(\frac{\Gamma(-1/2)}{4\Gamma(1/2)}-1\right).
\end{equation}
Notice the $\Gamma$-function $\Gamma(-1/2)$ is divergent. This divergence is artificial,
because
\begin{equation}
 \Gamma(-1/2) = \int_0^\infty dx x^{-3/2}\exp(-x).
\end{equation}
This divergence can be eliminated by a small but finite $\Gamma_0$ or a realistic
band structure. Therefore MR will be proportional to $\mu_H^2$. As we see in the
main paper, in FeAs$_2$, MR scales with $\mu_H^2$ over a wide temperature range,
whereas this is not the case for FeSb$_2$.

\end{document}